\documentclass[11pt]{article}
\usepackage{graphicx}
\usepackage{verbatim}
\usepackage{epsfig}

\date{}

\begin{document}

\title{The Genetic Programming Collaboration Network \\and its Communities}
\author{%
L. Luthi\thanks{Information Systems Department, University of Lausanne, Switzerland} \and 
M. Tomassini%
\addtocounter{footnote}{-1}%
\footnotemark
\thanks{Information Systems Department, University of Lausanne, Switzerland}
\and M. Giacobini%
\addtocounter{footnote}{-1}%
\footnotemark
\thanks{Dpt. of Animal Production Epidemiology and Ecology, University of Torino, Italy}
\and 
W. B. Langdon\thanks{Department of Computer Science, University of Essex, UK} 
}
\maketitle

\begin{abstract}
Useful information about scientific collaboration structures and patterns can
be inferred from computer databases of published papers.
The genetic programming bibliography is the most complete reference
of papers on GP\@.
In addition to locating publications, it contains coauthor and coeditor relationships 
from which a more complete picture of the field emerges. 
We treat these relationships as undirected small world graphs
whose study reveals the community structure of the GP
collaborative social network. 
Automatic analysis discovers new communities and
highlights new facets of them.
The investigation
reveals many similarities between GP and 
coauthorship networks in other scientific fields but also some subtle differences such as
a smaller central network component and a high clustering.
\end{abstract}

\section{Introduction}
\label{intro}

The genetic programming (GP) bibliography%
\footnote{http://www.cs.bham.ac.uk/$\sim$wbl/biblio/}, 
created and
maintained by one of us (WBL) and by S. Gustafson
contains
most of the GP papers.
As such, it is a rich
source of data that implicitly describes many aspects of the structure of the GP community.
Searching the bibliography and looking at the images%
\footnote{http://www.cs.bham.ac.uk/$\sim$wbl/biblio/gp-coauthors/}
provides a lot of useful information about the field and 
the people working on GP\@.
However, a deeper analysis of the data, that goes beyond the
mere pictorial aspect, provides a much more complete view.
The coauthorship data is a social network since
collaborating in a research study usually requires 
that the coauthors become
personally acquainted.
Thus, studying those ties, their structure, and their evolution
allows 
a better understanding of the factors that shape scientific collaboration.

We present a systematic study of the GP coauthorship data base
using methods and tools pertaining to complex networks and social network analysis.
Social network analysis (see \cite{social-nets-scott} for a survey), although it is an old discipline,
has recently received new impetus and tools from the field of complex networks
(see \cite{newman-03} for an excellent review).
This is mainly due to the relatively 
recent availability of large machine-readable databases such as the GP bibliography.
Social acquaintances involve psychological and other human aspects that are difficult to
quantify. However, as it has been done in other fields  \cite{barab-collab-02,grossman-02,newman-collab-01-1,newman-collab-01-2},
we use objective data such as
joint published work to stand for social bonds.
Since this must ignore subtler aspects of a collaboration relationship, 
it is obviously far from perfect as a social indicator,
yet it is still a good ``proxy'' for the network of social
relationships and can reveal several interesting facts and trends.

A preliminary investigation of the GP coauthorship network appears in \cite{TLGL-GPEM-07}. In the first part of this article we update this initial study using the
most recent data and adding the study of the influence of excluding co-edited proceedings
and books. In the second part we offer a new analysis of the finer community structure of the collaboration network. 
Similar studies have been performed in the last few years on several other collaboration networks in disciplines
such as physics, mathematics, medicine, biology, and computer science 
\cite{barab-collab-02,grossman-02,newman-collab-01-1,newman-collab-01-2}. 
A related investigation concerning the EC collaboration network
\cite{cotta-jjm} has appeared recently in popular form, but it does not take into account,
for example the
community structure of the network. 
\cite{cotta-jjm} deals with some of 
the same statistical features for the EC community at large
as we describe in detail here for GP\@.
The values reported by \cite{cotta-jjm} 
are in line with those found here for the GP field. 
Given that the intersection
between the GP researchers and general EC is likely to be rather large,
it would be interesting to
study how they are related to each other.

\section{The GP Collaboration Network}
\label{sec:collab}

We treat the genetic programming social network
as a graph where each node
is a GP researcher, i.e.\
someone who
has at least one entry in the bibliography. There is a connection
between two people if they have coauthored at least one paper, or
if they have coedited one or more book or proceedings.
As of the start of 2007,
there is a total of $N=2809$ connected nodes, 
i.e.\ authors that have at least one GP collaborator,
 and a total
of $5853$ edges (collaborations) in the GP coauthorship network. 
There are $367$ isolated vertices, which represent 
authors who have not collaborated with others 
to the extent of coauthoring a paper.
Isolated vertices are ignored in our graph statistics.
We have also excluded a single
paper with $108$ coauthors in a nuclear physics journal.
This is because
we consider it to be an anomalous entry 
that is not representative of typical collaborations
in our discipline. 

Due to the youth of GP,
the graph is relatively
small compared to some studied
collaboration networks \cite{barab-collab-02,grossman-02,newman-collab-01-1}. 
(Although some published studies have covered much smaller
and more specialised networks,
e.g.\ of only 50 people \cite{andrews:2003:JMLA}.)
The main disadvantage of studying a relatively small database is that,
like any statistical study,
more data allows deeper and more meaningful inferences to be drawn.
In particular,
studies 
of the form of the distributions 
(such as whether they follow exponential or power laws)
require a large amount of data.
The advantages include that the graph almost fully represents the
state of the whole GP community.
This allows reliable characterisation
of collaboration in the community. Also, the problems of 
multiple authors with the same name
(e.g. A. Smith),
outliers and different name spelling that plague the larger data sets, are unlikely and easy to spot in our data.

Although in many cases in our field
co-editing a book or proceedings volume does reflect personal acquaintance,
there are some large coeditorships which are not representative
and so may give a slanted view.
Therefore
in the following figures we present two kinds of statistics: those that include all joint publications and those in which co-edited conference proceedings and co-edited books are excluded
(but not their contents, of course).
Next we present and discuss some basic measures that characterise 
the GP collaboration network.

\begin{figure}
\centering
\epsfig{file=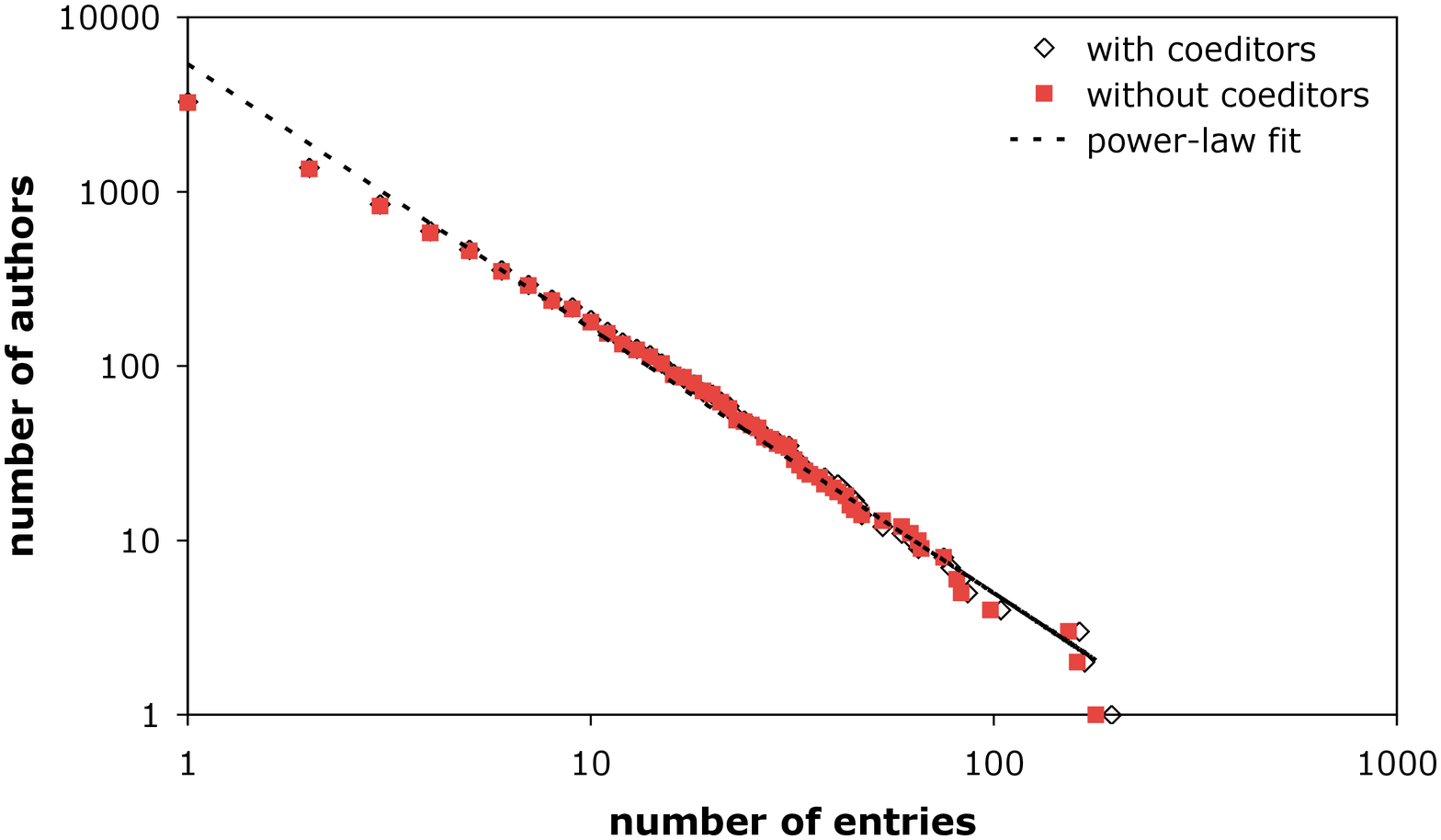,width=8cm,height=6.0cm}
\caption{Cumulative distribution of the number of entries per author. 
Log-log scale.
The straight line is the best mean-square fit
and shows the number of authors is $\propto k^{-2.5}$.
}
\label{noppa}
\end{figure}

\subsection{Number of Papers per Author}


The average number of papers per author is $3.16$ with co-edited books and proceedings and
it is $3.14$ without. 
The five most prolific authors are, in
decreasing order: J. Koza, R. Poli,
W. B. Langdon, W.~Banzhaf and C. Ryan. If we exclude proceedings' co-editors the
ranking remains unchanged.
Naturally the distribution of
the number of papers per author, $P(k)$,
has some scatter, particularly in the
tail of the distribution. Thus, we present in Figure \ref{noppa} the graph of the cumulative
distribution $P(k \ge n)$ which is smoother and allows the same inferences to be made. 
The curves are rather well fitted by a straight line, and thus the
distributions follow a power-law $P(k) \propto k^{-\gamma}$
with a calculated exponent $\gamma$ of $2.5$ for both of them. A power-law distribution with similar exponents has been
observed for analogous collaboration networks, e.g.\ $2.86$ for a biological publication
database (Medline), $3.41$ for a computer science database (NCSTRL), $2.4$ for mathematics, and
$2.1$ for a neuroscience papers database \cite{barab-collab-02,newman-collab-01-1}. 
A smaller exponent (in absolute value) means that the tail of the distribution is more
stretched towards high values of degree.

\subsection{Number of Collaborators per Author}
The average number of collaborators per author, i.e.\ the mean degree $\langle k \rangle$ of the
coauthorship graph, is $4.17$ with proceedings and $3.62$ without. 
This is close to the 
values reported by studies of
computer science, physics (excluding high energy physics) and Mathematics,
suggesting GP follows similar collaboration patterns 
to those disciplines.
However it is much less than found in high energy physics and medicine.
See Table~\ref{compar-tab}.
In order and including co-edited volumes,
the five authors that have the largest number of collaborators are: 
W.~Banzhaf, J.A.~Foster, P.~Nordin, W.B.~Langdon, U.-M.~O'Reilly.
Without co-edited books the ranking is:
P.~Nordin, W.~Banzhaf, J.~Daida, C.~Ryan and R.~Goodacre.
The five ``pairs'' that have the highest number of coauthored papers are, in
decreasing order both with or without co-edited proceedings: J. Koza--M. A. Keane, R. Poli--W.B. Langdon, J. Koza--D. Andre,
J. Koza--F. Bennet and F. Bennet--M.A. Keane. 
This shows that J. Koza's group
has been tightly collaborating for a long time, a conclusion that is confirmed
in the community study of section \ref{subcomm}. It is also evident that the W.B.
Langdon--R. Poli association has been an extremely productive one.

\begin{figure}[!ht]
\centering
\epsfig{file= 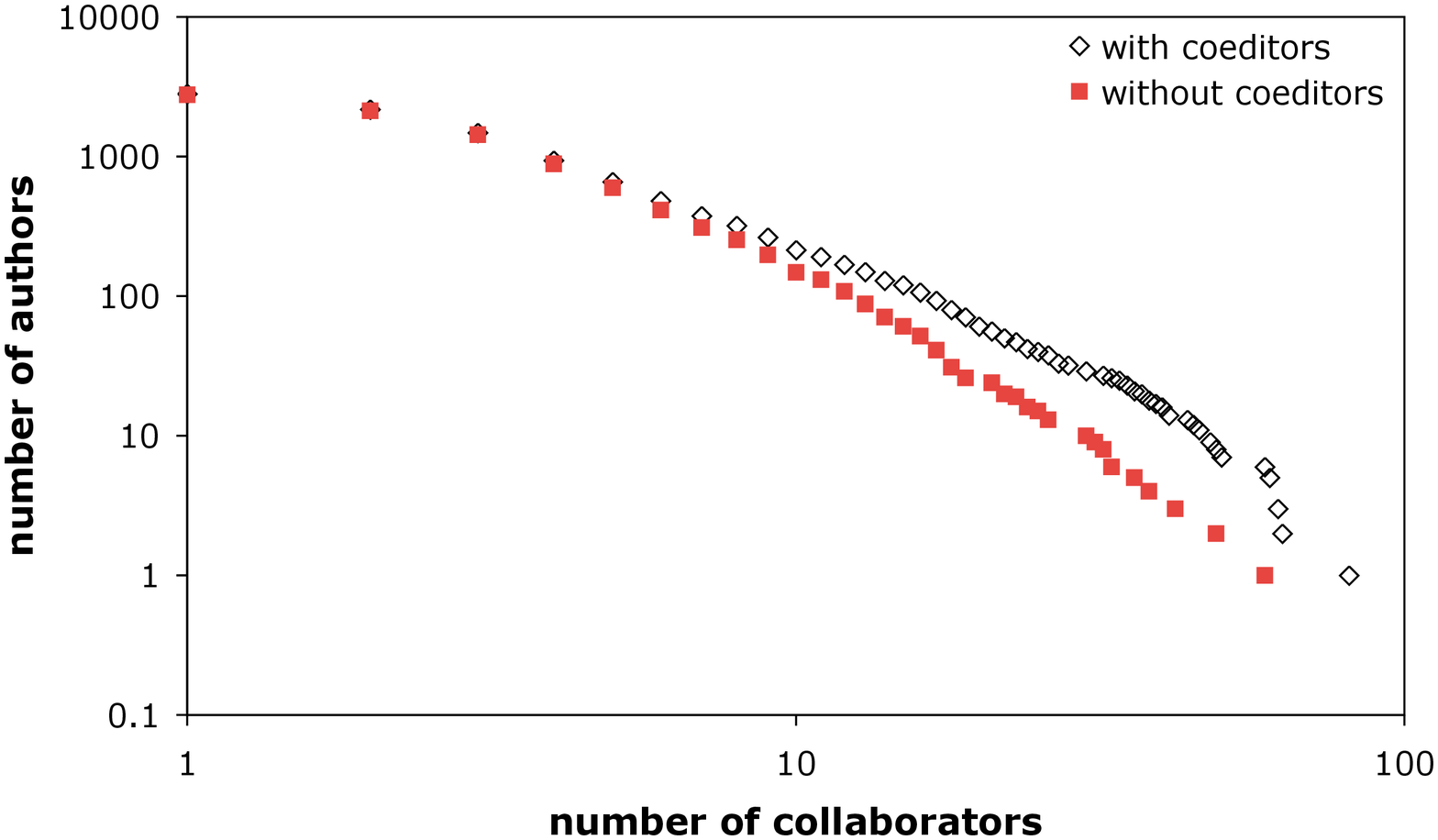,width=8cm,height=6cm}
\caption{Cumulative distribution of the number of authors with a given number of
collaborators. Logarithmic scale on both axes.}
\label{ncpa}
\end{figure}

Figure \ref{ncpa} shows the cumulative distributions of the number of collaborators. 
One sees that the distributions are  not  pure power-laws,
otherwise the points would approximately lie on a straight line. Rather, the distributions
shows a power-law regime
in the first part followed by an exponential decay in the tail. 
That is,
the whole network cannot be fitted by a power-law. This is 
quite common.
In fact,
several measured social networks do not follow a power-law degree distribution~%
\cite{am-scala-etc-2000,newman-collab-01-1} and are best fitted either by an exponential
degree distribution $P(k) \approx e^{-k/{\langle k \rangle}}$ or by an exponentially truncated power-law
of the type $P(k) \approx k^{-\gamma} e^{-k/k_c}$, where $k_c$ represents a critical connectivity
and $\langle k \rangle$ is the average degree.

\subsection{Number of Authors per Paper}
\begin{figure} [!ht]
\begin{center}
\includegraphics[width=8cm,height=6cm]{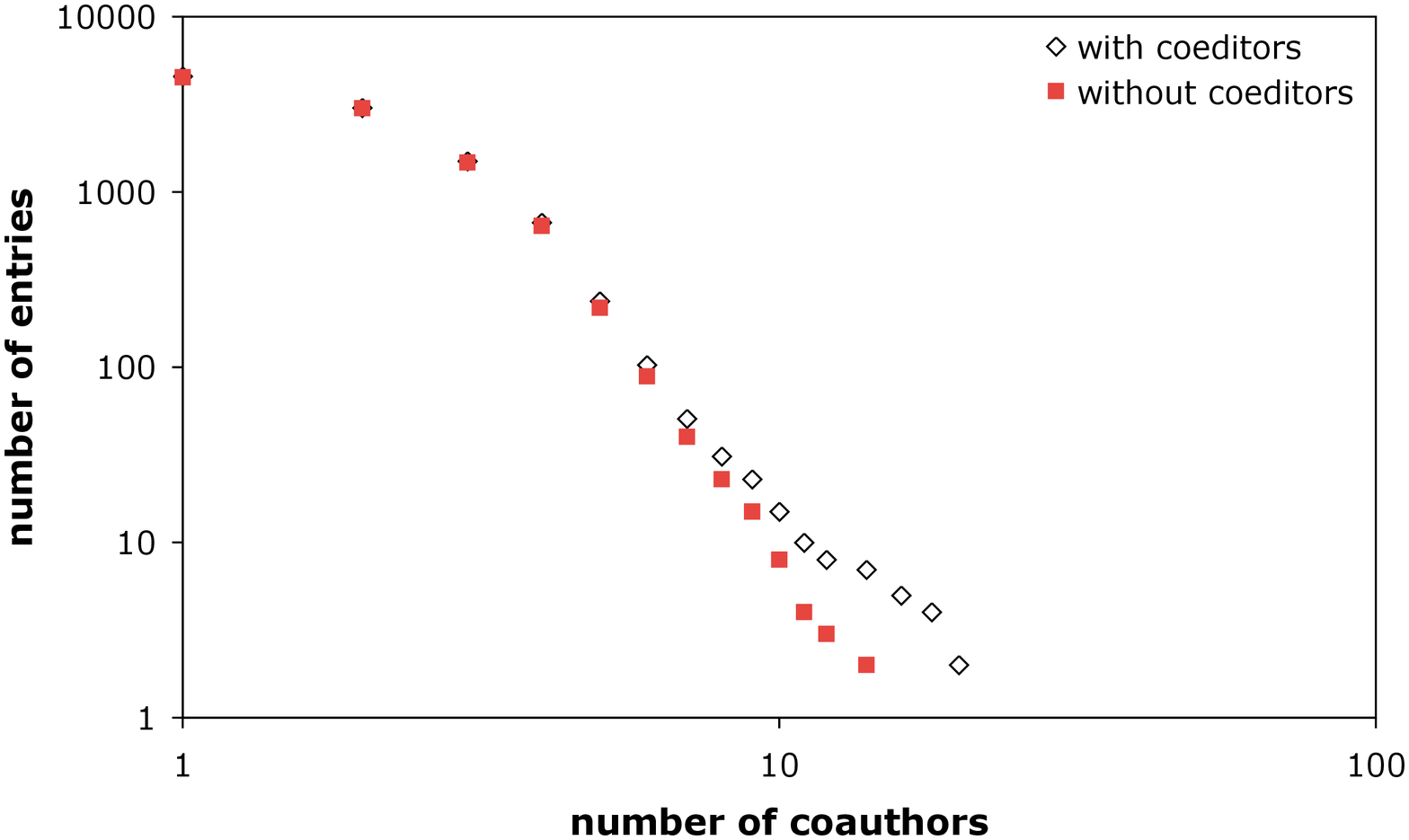}  
\caption{Cumulative distribution of the number of papers with a given number
of coauthors on log-log scales.}
\label{coapp}
\end{center}
\end{figure}

Figure \ref{coapp} shows the cumulative distribution of the number of papers written by
a given number of coauthors.
Here the distribution also has a tail that is longer than that of a 
Gaussian 
or exponential
distribution,
however it does not follow a power-law.
The average number of authors per paper is $2.25$ ($2.22$ without co-editors). From Table~\ref{compar-tab} we can see that
these figures are close to the equivalent ones for computer science (NCSTRL) 
and physics, while
Mathematics has a lower number of co-authors per paper. On the other hand, nuclear physics
stands out with an unusually high number of coauthors per paper.

From Figures \ref{ncpa} and \ref{coapp} one can see that the tails of the distribution
with co-editors are longer than without them. Thus, taking co-editorship into account
seems to rather artificially inflate the number of publications with many co-authors and,
by consequence, the number of collaborators that a person has. 

\subsection{Connected Components}
In the theory of Poisson random graphs there is a critical value of average degree
$ \langle k \rangle = 1$ above which
there is a sudden appearance of a 
\textit{giant component}.
This is so-called since
most vertices belong to it.
The other components
are smaller
and have an exponentially decreasing size distribution~\cite{newman-03}.
Although collaboration
graphs are not random, a similar phenomenon appears.
Including coeditors
there are $1025$ GP authors
in the giant component.
This is $36.5\%$ of the total graph.
If we exclude coediting proceedings etc.\
the size is $743$, representing the $26.9\%$ of the total. 
In the giant component 
the average number of collaborators per author is $5.83$ with co-editors and
$4.39$ without them. 

\begin{figure}[!ht]
\centering
\epsfig{file=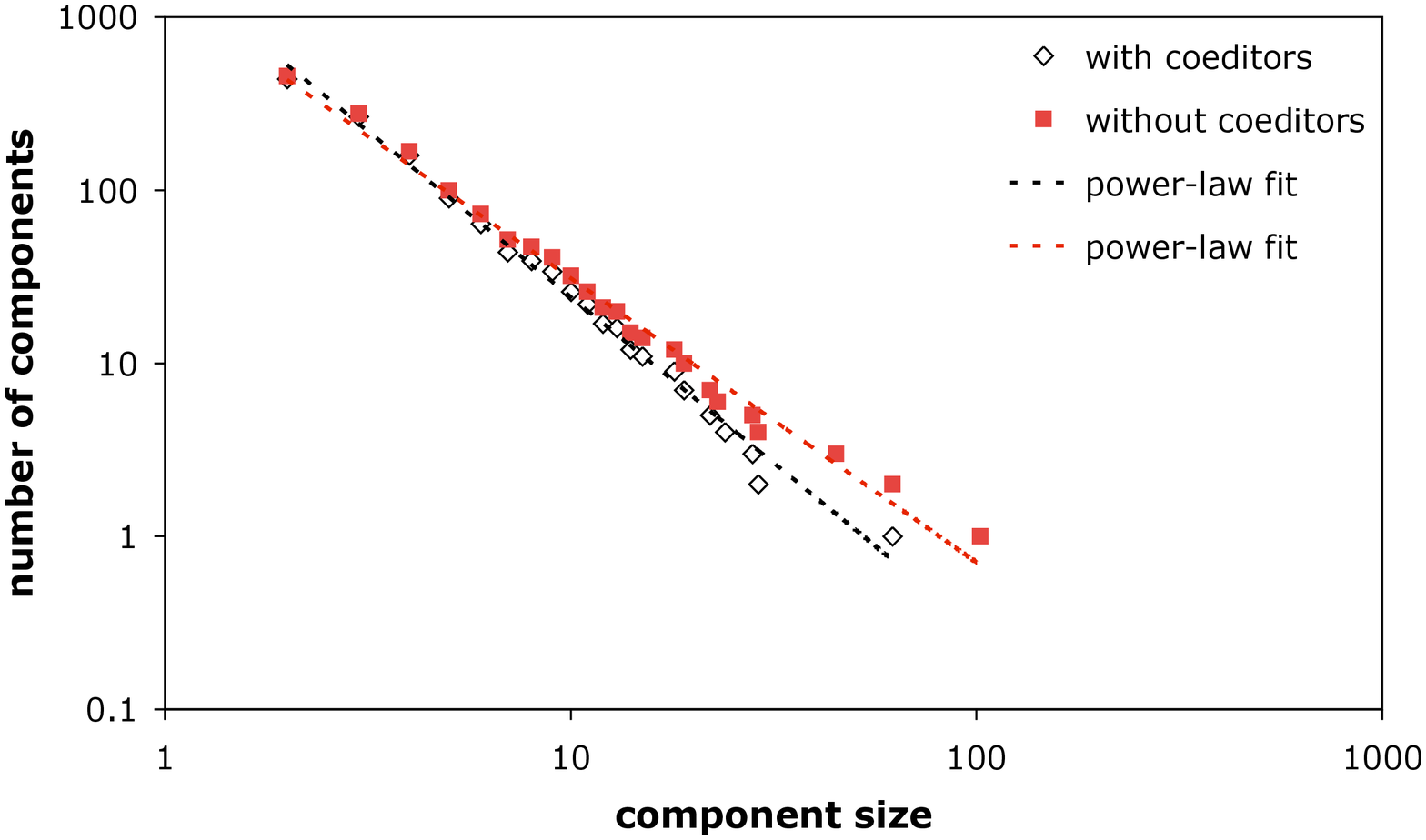,width=8cm,height=6cm}
\caption{Cumulative distributions of the number of connected
components in the collaboration graph by number of people.
Log-log scale.}
\label{ccd}
\end{figure}

The cumulative size distribution of the connected components
with and without co-editors are depicted in Figure \ref{ccd}.
Figure~\ref{ccd}
shows that the
probability density functions are well approximated by a power law with 
exponent of $2.9$ (excluding co-editors)
and $2.6$ (total). Since 
the other authors did not provide the
analogous data for their databases, we do not
know how our figures would compare with those for other coauthorship databases.

The existence of a big connected component has
a social meaning.
It suggests $36.5\%$ of GP researchers
are members of a single community, since those researchers are either directly connected
via a collaboration or they are close to each other in a way 
that will be made clear in section \ref{global}. The size of the giant component is notably
smaller
in the GP graph with respect to other measured coauthorship networks (see Table~\ref{compar-tab}).
This may be due to 
the comprehensive nature of the GP bibliography.
It captures work done by smaller groups which does not get into major journals,
whereas, perhaps, the other databases concentrate upon higher impact
outlets where work is heavily cited but at the expense of ignoring
less regarded authors.
This may artificially inflate the fraction of authors within their
giant component.
Alternatively it may be due to
the youth of the GP field, with many semi-isolated individuals
and groups starting research independently.

One should also consider that all collaboration networks are in a non-equilibrium state as they
are continuously evolving \cite{barab-collab-02}. 
Accordingly, as time goes by, one should observe small components progressively connecting themselves
to the large one. For example, in less than one year the size of the giant component including
co-editors has grown from 942 to 1025 nodes. This is due in part to a number of newcomers
collaborating with people already belonging to the giant component. The other part comes
from the absorption of a few disconnected small components into the giant one thanks to one or
more new collaborations.
This suggests that the size of the giant component has not
yet reached its ``steady-state'' value and it will continue to grow in relative size. Since we possess all the time-stamped data, it is
possible to study the evolution of this component, as well as several other indicators
from the beginning and up to the present days. This investigation is currently under way.

\subsection{Social GP Clusters}

The clustering coefficient of a node in a graph is the proportion of
its neighbouring nodes which 
are also neighbours of each other. The average clustering
coefficient
$\langle C \rangle$ is calculated across all nodes in the graph
 \cite{newman-03}. In other words, $\langle C \rangle$ is a simple statistical measure
of the amount of local structure that is present in a graph.
Most real-world networks, 
e.g.\
the  world wide web,
roads, electrical power transmission and
including the social networks that have been studied to date,
have a much larger clustering coefficient
than would be expected
of a random graph with the same number of vertices and edges. Social networks are
particularly clustered.
For example, the average clustering coefficient is
$\langle C \rangle=0.665$ for the GP collaboration graph including book co-editors, and it is $0.660$ without.
(We would expect $0.0015$ and $0.0013$ for the corresponding random graphs).
 In terms of scientific
collaborations, a high clustering coefficient means that people tend to collaborate in groups of three
or more.
This agrees with what we know of the GP field.
It may mean that two researchers that
collaborate independently with a third one may, in time, become acquainted and
so collaborate
themselves. 
Alternatively it might be due to collaborators
coming from the same institution. In all cases, a high value of $\langle C \rangle$ for a
social network is
an indication that collaborations are not made at random at all, and that social forces
and processes are at work in the network structure formation.

\begin{table*} [!ht]
\tiny
\begin{center}
\caption{Basic statistics for some scientific collaboration networks. GP1 is
the GP bibliography at the start of 2007,
including coedited books and proceedings. GP2 is the same but without coeditors. SPIRES is a data set of papers in high-energy physics. Medline is a database of articles on biomedical research. Mathematics comprises articles from
\textit{Mathematical Reviews}. NCSTRL is a database of preprints in computer science.
Physics has been assembled from papers posted on the Physics E-print Archive.
Details about these databases can be found in
\protect\cite{grossman-02,newman-collab-01-1,newman-collab-01-2}.
}
\vspace{0.3cm}
\begin{tabular}{|l  | c | c | c | c | c | c | c | }

\hline

   &  GP1 & GP2 & SPIRES &  Medline &  Mathematics & NCSTRL & Physics \\ \hline

Total number of papers & 4564 & 4504 & 66652 & 2163923 & 1600000 & 13169 & 98502 \\
Total number of authors & 2809 & 2765 & 56627 & 1520251 & 253339 & 11994 & 52909 \\
Average papers per author & 3.16 & 3.14 & 11.6 & 6.4&  7 & 2.55 & 5.1 \\
Average authors per paper & 2.25 & 2.22  & 8.96 & 3.754 & 1.5 & 2.22 & 2.53  \\
Average collaborators per author &  4.17 & 3.62  & 173 & 18.1 & 2.94 & 3.59 & 9.7 \\
Size of the giant component (\%) &  36.5 & 26.9 & 88.7  & 92.6 & 82.0 & 57.2 & 85.0 \\
Clustering coefficient & 0.665 & 0.660 & 0.726 & 0.066 & 0.15 & 0.496 & 0.43 \\ 
Average path length & 4.74 & 5.2  & 4.0  & 4.6 & 7.73 & 9.7 & 5.9 \\ \hline
\end{tabular}
\label{compar-tab}
\end{center}
\end{table*}

Table \ref{compar-tab} summarises
 the results of this section and compares them with those for some other collaboration networks. 
 Some of the entries in the table will be discussed in the following section.
Most GP statistics
are similar to those of the larger databases.
However
one notable difference, as we have already remarked, is the relative smallness of
the largest component.
The clustering is rather high, which shows that GP researchers know each
other quite well within the large component, and the community is rather homogeneous. 
In contrast, in
biology and medicine or mathematics,
where scientist from different sub-disciplines seldom
collaborate,
the clustering coefficient is lower.
Note also the high number of authors per paper, and especially the strikingly 
high number of collaborators
per author in the nuclear physics community (SPIRES). Clearly, nobody
can maintain an average of $173$ scientific partners on a first-hand acquaintance basis and thus this figure does not seem to be socially meaningful.

\section{Distances and Centrality}
\label{global}

A social network can be characterised by a number of measures that give an idea
of ``how far'' people are from each other, or how ``central'' they are with respect
to the whole community. These measures are well known in social network analysis.
Here we shall concentrate on \textit{average path length} and on \textit{betweenness centrality}.

\subsection{Average Path Length}
The average path 
length $L$ of a graph is the average value of the shortest paths between all of its pairs
of vertices.
In random graphs and
many real networks, such as the Internet, the World Wide Web 
and social networks,
the average path distance
scales as a logarithmic function O($\log N$) of the number of vertices $N$.
Such networks, if they also have a high clustering coefficient,
are known as \textit{small worlds} networks
\cite{Milgram1967Small}.
Since, even for very large graphs,
any two nodes in a small world network are only a few steps apart.
In contrast in regular lattices,
two nodes are O($N^\frac{1}{D}$) apart.
(Where $D$ is the lattice's dimensionality.
For example, for a square lattice $L\leq \frac{2}{3}N^\frac{1}{2}$).
The average path length of the giant component of the GP collaboration graph including
coeditors is
$4.74$ (it is $5.2$ without coeditors). The longest among all the shortest paths
(known as the diameter)
is $12$ ($14$ without coeditors).
Thus, unsurprisingly, the GP community, as far as its ``core'' component is concerned,
 is indeed a small world and is characterised by values that are typical of these kinds of network (see
 Table \ref{compar-tab}). Being a small world means that information may circulate
quickly and collaborations are easier to set up. 
These are clearly advantageous for a research
community.
The connected components following the largest one are themselves small worlds. 
We expect over time
some of them will merge  with
the largest component.
(For this to happen, only a single new collaboration between two scientists each belonging to one of the components is needed.)

\subsection{Betweenness}
The betweenness $b(v)$ of a vertex $v$ is the total number of shortest paths between all
possible pairs of vertices that pass through this vertex. Nodes that have a high betweenness
potentially have more influence, 
i.e.\ they are more central in the network, in that there is more ``traffic'' that goes through them. 
The first five authors in terms of betweenness in the network 
(including co-editors and in decreasing order)
are: W. Banzhaf, H.~Iba, U.-M. O'Reilly, H.~de~Garis and W. B. Langdon. 
W.~Banzhaf is also the researcher that has the highest number of different collaborators. Without co-editors the ranking is: 
W. B. Langdon, U.-M. O'Reilly,  W.~Banzhaf,
M. Tomassini and P. Nordin. People who have
a large value of betweenness play the role of intermediaries or ``brokers'' in a social sense.

\begin{figure*} [!ht]
\centering
\epsfig{file=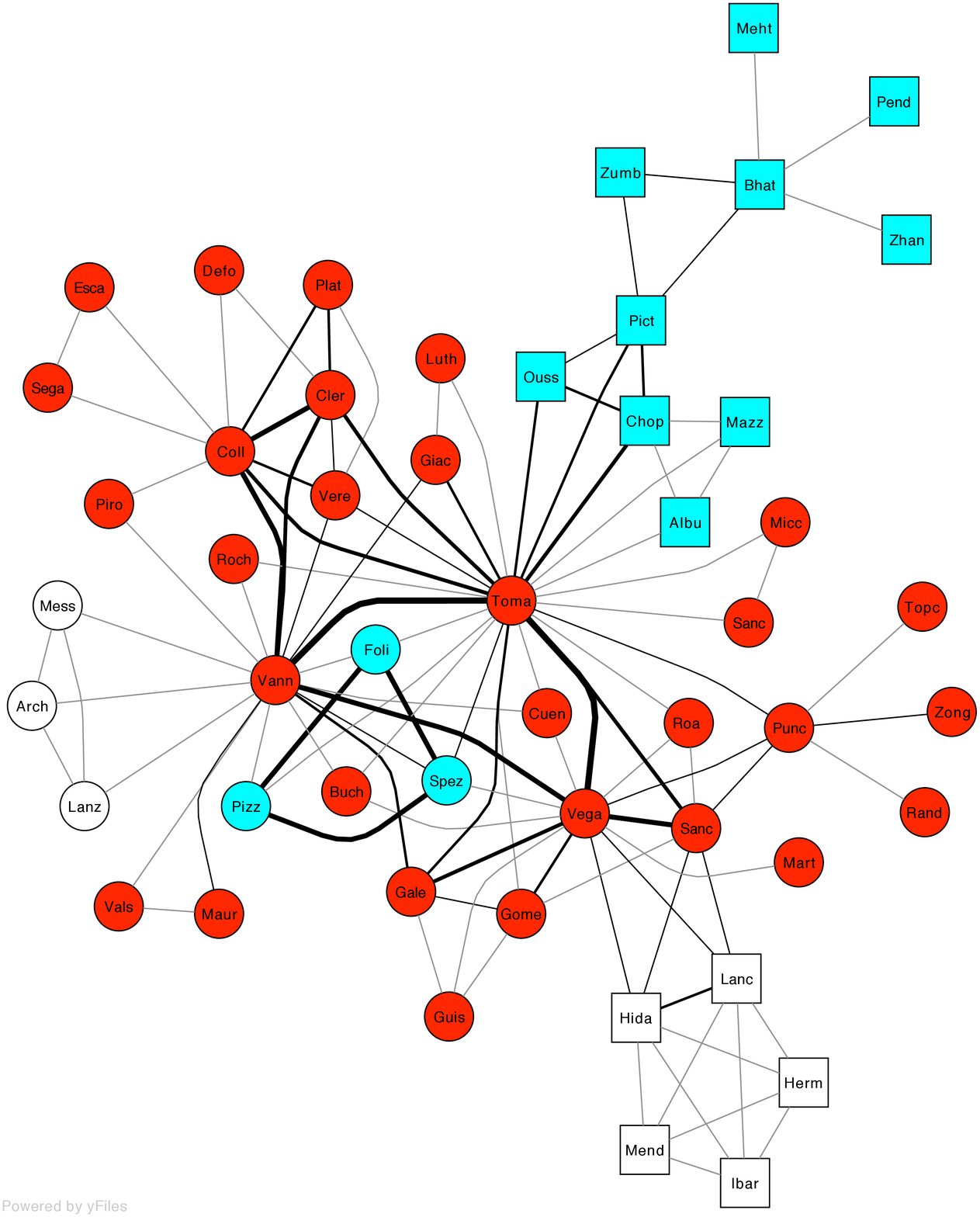,width=13cm,height=13cm}
\caption{One of the communities belonging to the main network component. The thickness of
the links gives an indication of the number of co-authored papers. The largest thickness
indicates more than 16 coauthored works. The thinnest link (light gray) stands for a single
collaboration. The different symbols and colours represent sub-communities of the 
illustrated community.}
\label{commmt}
\end{figure*}

\subsection{Non-random collaborations between\\ directly connected authors} 

Most technological and biological networks are disassortive in that they have  negative correlation, meaning that high-degree vertices are preferentially connected to low-degree vertices.
However most
measured social networks are assortative,
meaning highly connected nodes tend to be connected
with other highly connected nodes \cite{newman-03}. 
The GP collaboration network confirms this general observation with a correlation coefficient of $+0.15$ for the giant component, and
$+0.30$ for the whole graph (including coeditors and excluding the single physicist's paper).
%
%
These are close to
the coefficients
observed for other social networks
(specifically 
0.127 for Medline 
 and
0.120 for Mathematics 
\cite{newman-assort-02}).

\section{Communities in the Giant Component}
\label{subcomm}

All the researchers belonging to the largest component of the network can be said to
form part of the GP community at large,
in the sense that they are all only a few steps away from any
other member of the community. However, we know from direct experience that some groups 
of GPers are more
closely connected between themselves than with other people. 
In other words, they belong to
what one might call a group or a tighter community within the global one. 
It is not easy to give a rigorous quantitative
definition of a community within a network. For our purposes
a community can be seen as a set of highly connected vertices having few connections with vertices belonging to other communities. 
In the analysis of social networks, several algorithms that attempt to split a network into communities have been proposed. 
We used Newman's method \cite{newman-2004-69}, which is based on a measure of the fraction
of edges that fall within communities minus the expected value of the same quantity if edges
fall at random without regard for the community structure.

\begin{figure*}[!ht]
\centering
\epsfig{file=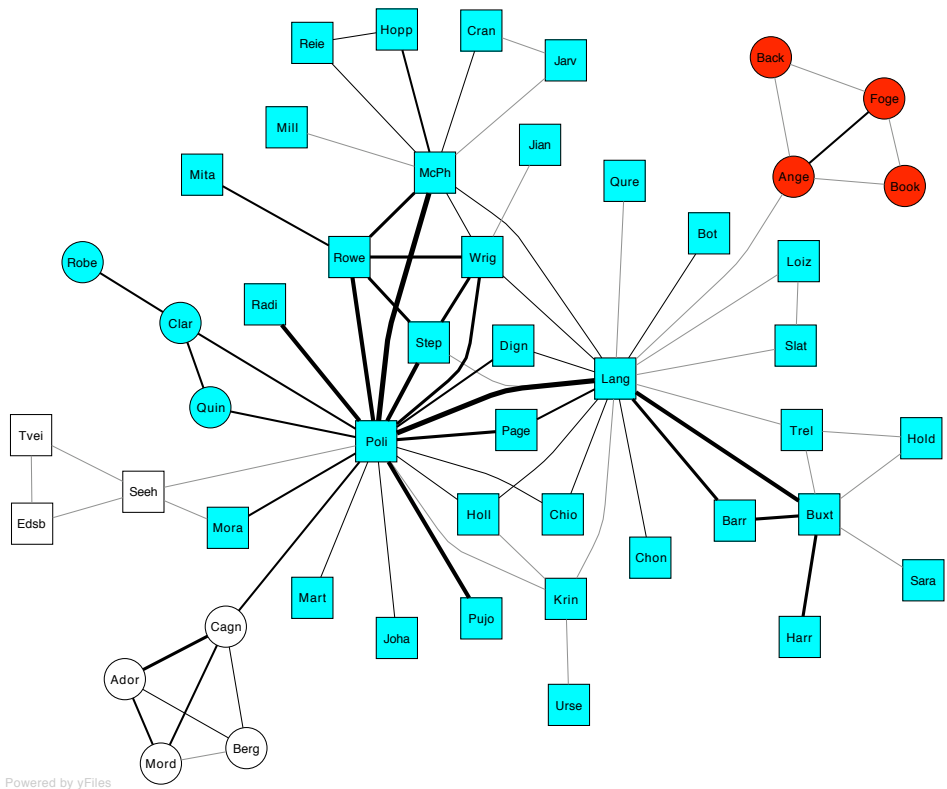,height=11cm}
\caption{Another community belonging to the main network component. The thickness of
the links gives an indication of the number of co-authored papers. The largest thickness
indicates more than 16 coauthored works. The thinnest link (light gray) stands for a single
collaboration. The different symbols and colours represent sub-communities.}
\label{commwbl}
\end{figure*}

Since the GP bibliography contains the number of papers that any two collaborators have published
together, it is possible to go a step further than just saying that two people have
coauthored at least a paper, and give a measure of the \textit{intensity} of the collaboration. 
We
use the number of papers that two given authors have in common as a measure of the strength of their collaboration.
Newman \cite{newman-collab-01-2} has proposed a more refined measure which takes into account the actual number
of coauthors of each paper. 
However this is more complicated than we need,
instead
we ignore the total number of coauthors for each paper.
Our measure of collaboration strength is used
in our communities algorithm 
to highlight groups of researchers that have collaborated strongly
with the aim of uncovering the
stability of the scientific relationship. We have also excluded coedited
proceedings, books, etc., as we have already seen that these might sometimes represent spurious
collaboration relationships.


The results of running the algorithm on the subgraph represented by the 
largest connected component
are qualitatively surprisingly close to what one would expect, 
given our knowledge of the GP field.
The
advantage is that the analysis makes them explicit and uncovers a number of other relationships
that would be difficult to infer without an explicit study of the raw data. As an example of
the about $25$ communities that the algorithm discovers, Figure \ref{commmt} shows the structure
of the groups around one of us (``Toma''). 
If we now consider this community as an isolated subgraph and apply again Newman's algorithm to it,
we obtain the groups highlighted by different symbols and colours in the figure. 
%
Thus, the groups correspond to sub-communities within the main community. 
The thickness of the links represents
the intensity of the relationship. It is easy to recognise a ``hard core'' of collaborating
researchers strongly connected to ``Toma'' forming triads and higher polygons of order four and
five. The strong triangle (``Foli'', ``Pizz'', ``Spez'') is relatively loosely connected to the rest, showing that
these researchers belong to the community but 
often collaborate between themselves.
It is also possible to discern institutional and geographical components in the community. For example, most of the upper right part of the figure through the node ``Chop'' comprises
researchers essentially belonging to the University of Geneva, which is close to the University
of Lausanne, to which ``Toma'' belongs. However, geographical closeness is not the key
factor in the other groups which belong to Universities in France, Italy, Spain, and the US\@.
We might conjecture that many collaborations start locally at the same or at close institutions
and then they spread through people being introduced
to others via a common acquaintance, or  through  people physically moving or
visiting other institutions. This is the case in the figure, where ``Vann'', ''Chop'', and
``Vega'' among others have played the role of ``bridges'' between different institutions and
across countries.

As a second illustration, let us look at Figure \ref{commwbl} which is the community
that revolves around one of us (``Lang'') and ``Poli''. In contrast to the previous case,
one can see that the graph structure is more ``star-like'', with two large directly
connected big hubs (``Lang'' and ``Poli'') who have about 70 co-authored papers, and three other highly connected nodes (``Buxt'',``McPh'', ``Rowe'') which are strongly connected to one of the main hubs but not to both.
It is interesting to observe the role of ``McPh'' who, like ``Vega'' in 
the previous community
(cf.\ Figure~\ref{commmt}),
plays a bridging role,
this time between some UK and some American researchers.
We can also recognise a strong "theory-oriented" group,
which is almost a clique in the graph, formed
by (``McPh'',``Poli'',``Rowe'', ``Steph'', ``Wrig''). There is also another bridge formed
by ``Cagn'' from UK to Italy,
again due to a long-standing collaboration and friendship.
The small cliques or almost cliques at the periphery of the figure 
essentially represent
people that have worked at the same institution in either Italy or the US.

The discussion above, motivated by our belonging to the mentioned communities, and thus by
our direct human knowledge about them, should be enough to get an impression of the many useful
observations that one can make on the communities that interlock in the main network
component.
There are of course several other large well known and interesting communities in the network but unfortunately we cannot
describe them here for reasons of space.


\section{Conclusions}
\label{concl}

In sections~\ref{sec:collab} and~\ref{global}
we characterised the genetic programming (GP) coauthorship graph 
using a number
of local and global statistics. We extended and updated the
findings presented in \cite{TLGL-GPEM-07} by studying the influence of coedited
volumes and by using the latest data available. 
Section~\ref{global}
showed the GP field to be
highly clustering
and that the GP coauthorship network has a small mean path length.
Together these suggest that,
at least for the core, 
GP is indeed a ``small world''.
We also found,
compared with other published collaboration networks,
that the
fraction of GP authors connected by coauthorship
is a relatively small fraction of all GP authors.

Section~\ref{subcomm} is a
study of the community structure of GP\@.
It uses a more precise definition of collaboration, 
which takes into account
the intensity of the relationship.
This uncovers many 
groups of tightly interacting researchers. 
From the detailed study of two of the
communities we have drawn inferences
about the pivotal role of some researchers or groups of 
researchers in promoting collaborations within and between academic institutions. Adding
our human knowledge about geographical location and personal acquaintance, 
allows some
conjectures to be drawn about the way in 
which different continents and countries
collaborate on research projects.

It should be obvious that the present data driven approach to social network analysis can only
provide some answers but not all of them. Algorithms and data cannot take into account human aspects such as
 friendship in scientific
collaboration. 
While these may be buried in the sea of numbers
they will never appear explicitly from such analyses. Nevertheless, we feel that
our results are interesting and useful in the way that they characterise our community.

There is another aspect of the collaboration graph that would be revealing: the
analysis of its development over the years.
Indeed, since each paper has a date of publication,
we possess all the data that are needed for such
an investigation.
This would allow the detailed study of
how the network has grown to its present size and structure from the beginning
and might give hints as to its future progress.
This extension is
currently under investigation.


%
\bibliographystyle{abbrv}
\bibliography{refs}  
%
%
%
\end{document}